\UseRawInputEncoding
\documentclass[runningheads,openbib]{llncs}

\usepackage{xcolor}
\usepackage{listings}
\usepackage{inconsolata}
\usepackage{relsize}
\usepackage{mathpartir}
\usepackage{graphicx}
\usepackage{xspace}
\usepackage{xurl}
\usepackage{caption}
\usepackage{bbding}
\usepackage[paperheight=235mm, paperwidth=155mm,textwidth=12.2cm,textheight=19.3cm]{geometry}

\usepackage[bookmarks,unicode,colorlinks=true,linkcolor=blue,citecolor=blue,urlcolor=blue]{hyperref}

\newcommand{\Section}[1]{\vspace{-0.5ex}\section{#1}}
\newcommand{\SubSection}[1]{\vspace{-1.0ex}\subsection{#1}}
\newcommand{\Paragraph}[1]{\vspace{-1.0ex}\subsubsection{#1}} 

\newenvironment{Figure}{
  \begin{figure}[tb]
} {
  \end{figure}
}

\newcommand{\MVP}[0]{\textsf{MVP}\xspace}
\newcommand{\solidity}[0]{\textsf{Solidity}\xspace}

\usepackage[charter]{mathdesign}

\lstdefinestyle{MoveStyle}{
  basicstyle=\ttfamily,
  keywordstyle=\color{black}, 
  escapechar=@, 
  breaklines=true,
}

\lstdefinelanguage{Move}{
  morekeywords={
    abort,
    aborts_if,
    acquires,
    address,
    as,
    assert,
    assume,
    borrow_global,
    borrow_global_mut,
    break,
    const,
    continue,
    copy,
    copyable,
    define,
    drop,
    else,
    ensures,
    exists,
    false,
    forall,
    friend,
    fun,
    global,
    has,
    havoc,
    if,
    include,
    invariant,
    key,
    let,
    loop,
    modifies,
    module,
    move,
    move_from,
    move_to,
    mut,
    native,
    num,
    old,
    onabort,
    pragma,
    public,
    requires,
    resource,
    return,
    schema,
    script,
    signer,
    spec,
    store,
    struct,
    true,
    u8,
    u64,
    u128,
    update,
    use,
    with,
    where,
    while},
  sensitive=true,
  morecomment=[l]{//},
  morecomment=[s]{/*}{*/},
}

\lstMakeShortInline[language=Move,style=MoveStyle]|

\lstnewenvironment{Move}{
  \lstset{
    language=Move,
    style=MoveStyle,
    basicstyle=\relsize{-1}\ttfamily,
    keywordstyle=\color{blue},
  }
}{
}

\lstnewenvironment{MoveBox}{
  \lstset{
    language=Move,
    style=MoveStyle,
    basicstyle=\relsize{-1}\ttfamily,
    keywordstyle=\color{blue},
  }
}{
}

\lstnewenvironment{MoveBoxNumbered}{
  \lstset{
    language=Move,
    style=MoveStyle,
    basicstyle=\relsize{-1}\ttfamily,
    keywordstyle=\color{blue},
    numbers=left,
    numberstyle=\scriptsize\color{gray},
  }
}{
}

\lstnewenvironment{MoveDiag}{
  \lstset{
    style=MoveStyle,
    basicstyle=\relsize{-2}\ttfamily,
  }
}{
}

\begin{document}

\author{
  David Dill \and Wolfgang Grieskamp(\Envelope)\thanks{\mailname~\email{wgrieskamp@gmail.com}} \and \\ Junkil
  Park \and Shaz Qadeer \and Meng Xu \and Emma Zhong
}

\institute{Novi Research, Meta Platforms, Menlo Park, USA}
\authorrunning{D. Dill, W. Grieskamp et. al.}

\title{Fast and Reliable Formal Verification of Smart Contracts with the Move
  Prover\thanks{To appear in TACAS'22. This is the extended version with
    appendices.}}

\maketitle
\begin{abstract}
  The Move Prover (\MVP) is a formal verifier for smart contracts written in the
  Move programming language. \MVP has an expressive specification language, and
  is fast and reliable enough that it can be run routinely by developers and in
  integration testing.  Besides the simplicity of smart contracts and the Move
  language, three implementation approaches are responsible for the practicality
  of \MVP: (1) an alias-free memory model, (2) fine-grained invariant checking,
  and (3) monomorphization.  The entirety of the Move code for the Diem
  blockchain has been extensively specified and can be completely verified by
  \MVP in a few minutes. Changes in the Diem framework must be successfully
  verified before being integrated into the open source repository on GitHub.
\end{abstract}

\Section{Introduction}

The Move Prover (\MVP) is a formal verification tool for smart contracts that
intends to be used routinely during code development.  The verification finishes
fast and predictably, making the experience of running \MVP similar to the
experience of running compilers, linters, type checkers, and other development
tools.  Building a fast verifier is non-trivial, and in this paper, we would
like to share the most important engineering and architectural decisions that
have made this possible.

One factor that makes verification easier is applying it to smart contracts.
Smart contracts are easier to verify than conventional software for at least
three reasons: 1) they are small in code size, 2) they execute in a
well-defined, isolated environment, and 3) their computations are typically
sequential, deterministic, and have minimal interactions with the environment
(e.g., no explicit I/O operations).  At the same time, formal verification is
more appealing to the advocates for smart contracts because of the large
financial and regulatory risks that smart contracts may entail if misbehaved, as
evidenced by large losses that have occurred
already~\cite{CONTRACT_VERIFICATION,hacks-on-smart-contracts,hacks-on-compound}.

The other crucial factor to the success of \MVP is a tight coupling with the
\emph{Move} programming language~\cite{MOVE_LANG}.  Move is developed as part of
the Diem blockchain~\cite{DIEM} and is designed to be used with formal
verification from day one.  Move is currently co-evolving with \MVP.  The
language supports specifying pre-, post-, and aborts conditions of functions, as
well as invariants over data structures and over the content of the global
persistent memory (i.e., the state of the blockchain).  One feature that makes
verification harder is that quantification can be used
freely in specifications.

Despite this specification richness, \MVP is capable of verifying the full Move
implementation of the Diem blockchain (called the Diem
framework~\cite{DIEM_FRAMEWORK}) in a few minutes.  The framework provides
functionality for managing accounts and their interactions, including multiple
currencies, account roles, and rules for transactions.  It consists of about
8,800 lines of Move code and 6,500 lines of specifications (including comments
for both), which shows that the framework is extensively specified.  More
importantly, \emph{verification is fully automated and runs continuously with
  unit and integration tests}, which we consider a testament to the practicality
of the approach.  Running the prover in integration tests requires more than
speed: it requires reliability, because tests that work sometimes and fail or
time out other times are unacceptable in that context.

\MVP is a substantial and evolving piece of software that has been tuned and
optimized in many ways.  As a result, it is not easy to define exactly what
implementation decisions lead to fast and reliable performance.  However, we can
at least identify three major ideas that resulted in dramatic improvements in
speed and reliability since the description of an early prototype of
\MVP~\cite{MOVE_PROVER}:
\begin{itemize}
\item an \emph{alias-free memory model} based on Move's semantics, which are
  similar to the Rust programming language;
\item \emph{fine-grained invariant checking}, which ensures that invariants hold
  at every state, except when developer explicitly suspends them; and
\item monomorphization, which instantiates type parameters in Move's generic
  structures, functions, and specification properties.
\end{itemize}
The combined effect of all these improvements transformed a tool that worked,
but often exhibited frustrating, sometimes random~\cite{BUTTERFLY}, timeouts on
complex and especially on erroneous specifications, to a tool that almost always
completes in less than 30 seconds.  In addition, there have been many other
improvements, including a more expressive specification language, reducing false
positives, and error reporting.

The remainder of the paper first introduces the Move language and how \MVP is
used with it, then discusses the design of \MVP and the three main optimizations
above.  There is also an appendix that describes injection of function
specifications.


\Section{Move and the Prover}

Move was developed for the Diem blockchain \cite{DIEM}, but its design is not
specific to blockchains.  A Move execution consists of a sequence of updates
evolving a \emph{global persistent memory state}, which we just call the
\emph{(global) memory}.  Similar to other blockchains, updates are a series of
atomic transactions.  All runtime errors result in a transaction abort, which
does not change the blockchain state except to transfer some currency (``gas'')
from the account that sent the transaction to pay for cost of executing the
transaction.

The global memory is organized as a collection of resources, described by Move
structures (data types). A resource in memory is indexed by a pair of a type and
an address (for example the address of a user account). For instance, the
expression |exists<Coin<USD>>(addr)| will be true if there is a value of type
|Coin<USD>| stored at |addr|. As seen in this example, Move uses type generics,
and working with generic functions and types is rather idiomatic for Move.

A Move application consists of a set of \emph{transaction scripts}. Each
script defines a Move function with input parameters but no output
parameters.  This function updates the global memory and may emit
events. The execution of this function can abort because of an abort
instruction or implicitly because of a runtime error such as an out-of-bounds
vector index.

\Paragraph{Programming in Move}

\begin{Figure}
\caption{\label{fig:AccountDef} Account Example Program}
\begin{MoveBox}
module Account {
  struct Account has key {
    balance: u64,
  }

  fun withdraw(account: address, amount: u64) acquires Account {
    let balance = &mut borrow_global_mut<Account>(account).balance;
    assert(*balance >= amount, Errors::limit_exceeded());
    *balance = *balance - amount;
  }

  fun deposit(account: address, amount: u64) acquires Account {
    let balance = &mut borrow_global_mut<Account>(account).balance;
    assert(*balance <= Limits::max_u64() - amount, Errors::limit_exceeded());
    *balance = *balance + amount;
  }

  public(script) fun transfer(from: &signer, to: address, amount: u64)
  acquires Account {
    assert(Signer::address_of(from) != to, Errors::invalid_argument());
    withdraw(Signer::address_of(from), amount);
    deposit(to, amount);
  }
}
\end{MoveBox}
\end{Figure}

\noindent In Move, one defines transactions via \emph{script functions} which
take a set of parameters.  Those functions can call other functions. Script and
regular functions are encapsulated in \emph{modules}. Move modules are also the
place where structs are defined. An illustration of a Move contract is given in
Fig.~\ref{fig:AccountDef} (for a more complete description see the Move
Book~\cite{MOVE_LANG}). The example is a simple account which holds a balance in
the struct |Account|, and offers the script function |transfer| to manipulate
this resource.  Scripts generally have \emph{signer} arguments, which are tokens
which represent an account address that has been authenticated by a
cryptographic signature.  The |assert| statement in the example causes a Move
transaction to abort execution if the condition is not met. Notice that Move,
similar as Rust, supports references (as in |&signer|) and mutable references
(as in |&mut T|).  However, references cannot be part of structs stored in
global memory.


\Paragraph{Specifying in Move}

\begin{Figure}
\caption{\label{fig:AccountSpec} Account Example Specification}
\begin{MoveBox}
module Account {
  spec transfer {
    let from_addr = Signer::address_of(from);
    aborts_if from_addr == to;
    aborts_if bal(from_addr) < amount;
    aborts_if bal(to) + amount > Limits::max_u64();
    ensures bal(from_addr) == old(bal(from_addr)) - amount;
    ensures bal(to) == old(bal(to)) + amount;
  }

  spec fun bal(acc: address): u64 {
    global<Account>(acc).balance
  }

  invariant forall acc: address where exists<Account>(acc):
    bal(acc) >= AccountLimits::min_balance();

  invariant update forall acc: address where exists<Account>(acc):
    old(bal(acc)) - bal(acc) <= AccountLimits::max_decrease();
}
\end{MoveBox}
\end{Figure}

\noindent The specification language supports {\em Design By Contract}
\cite{DESIGN_BY_CONTRACT}. Developers can provide pre and post conditions for
functions, which include conditions over parameters and global
memory. Developers can also provide invariants over data structures, as well as
the contents of the global memory.  Universal and existential quantification
over bounded domains, such as like the indices of a vector, as well as
effectively unbounded domains, such as memory addresses and integers, are
supported.  Quantifiers make the verification problem undecidable and cause
difficulties with timeouts.  However, in practice, we notice that quantifiers
have the advantage of allowing more direct formalization of many properties,
which increases the clarity of specifications.

Fig.~\ref{fig:AccountSpec} illustrates the specification language by extending
the account example in Fig.~\ref{fig:AccountDef} (for the definition of the
specification language see \cite{MOVE_SPEC_LANG_DEF}). This adds the
specification of the |transfer| function, a helper function |bal| for use in
specs, and two global memory invariants. The first invariant states that a
balance can never drop underneath a certain minimum. The second invariant refers
to an update of global memory with pre and post state: the balance on an account
can never decrease in one step more than a certain amount.  Note that while the
Move programming language has only unsigned integers, the specification language
uses arbitrary precision signed integers, making it convenient to specify
something like |x + y <= limit|, without the complication of arithmetic
overflow.

Specifications for the |withdraw| and |deposit| functions have been omitted in
this example.  \MVP supports omitting specs for non-recursive functions, in
which case they are treated as being inlined at caller site.




\Paragraph{Running the Prover}
\label{sec:RunningProver}

\MVP is fully automatic, like a type checker or linter, and is
expected to finish in a reasonable time, so it can be integrated in
the regular development workflow. Running \MVP on the |module Account| produces
multiple errors. The first is this one:

\begin{MoveDiag}
error: abort not covered by any of the `aborts_if` clauses
   -- account.move:24:3
   |
13 |       let balance = &mut borrow_global_mut<Account>(account).balance;
   |                          ----------------- abort happened here
   |
   =     at account.move:18: transfer
   =         from = signer{0x18be}
   =         to = 0x18bf
   =         amount = 147u8
   =     at ...
\end{MoveDiag}

\noindent \MVP detected that an implicit abort condition is missing in the
specification of the |withdraw| function. It prints the context of the error, as
well as an \emph{execution trace} which leads to the error. Values of variable
assignments from the counterexample found by the SMT solver are printed together
with the execution trace. Logically, the counterexample presents an
assignment to variables where the program fails to meet the specification. In
general, \MVP attempts to produce readable diagnostics for Move developers
without the need of understanding any internals of the prover.

There are more verification errors in this example, related to the global
invariants: the code makes no attempt to respect the limits in |min_balance()|
and |max_decrease()|.  The problem can be fixed by adding more |assert|
statements to check that the limits are met (see Appendix~\ref{sec:CorrectedAccount}).

The programs and specifications \MVP deals with are much larger than
this example. The conditions under which a transaction
in the Diem framework can abort typically involve dozens of individual predicates,
stemming from other functions called by this transaction. Moreover, there are
hundreds of memory invariants specified, encoding access control and other
requirements for the Diem blockchain.


\Section{Move Prover Design}

\begin{Figure}
  \centering
  \caption{Move Prover Architecture}
  \label{fig:Arch}
  \includegraphics[trim=0 250 0 0, width=\textwidth]{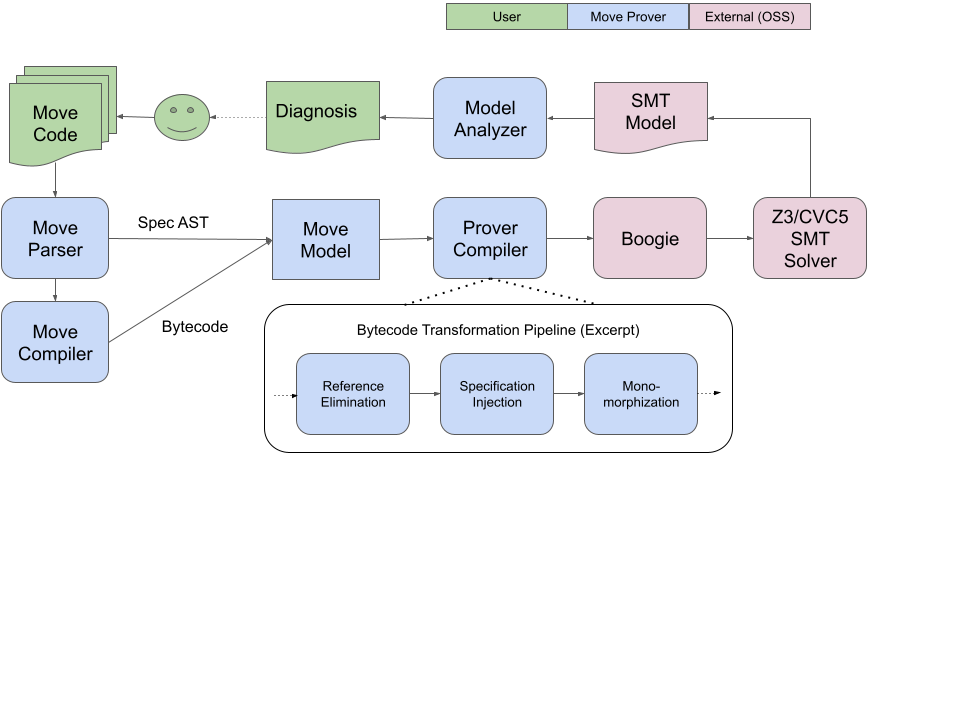}
\end{Figure}

The architecture of \MVP is illustrated in Fig.~\ref{fig:Arch}. Move code
(containing specifications) is given as input to the tool chain, which
produces two artifacts: an abstract syntax tree (AST) of the specifications,
and the generated bytecode.  The \emph{Move Model} merges both bytecode and
specifications, as well as other metadata from the original code, into a
unified object model which is input to the remaining tool chain.

The next phase is the actual \emph{Prover Compiler}, which is a
pipeline of bytecode transformations. We focus on the
transformations shown (Reference Elimination, Specification Injection, and
Monomorphization).
The Prover uses a modified version of the Move VM bytecode as an intermediate
representation for these transformations, but, for clarity,
we describe the transformations at the Move source level.

The transformed bytecode is next compiled into the Boogie intermediate
verification language \cite{BOOGIE}. Boogie supports an imperative programming
model which is well suited for the encoding of the transformed Move code. Boogie
in turn can translate to multiple SMT solver backends, namely Z3 \cite{Z3} and
CVC5 \cite{CVC}; the default choice for the Move prover is currently Z3.



\SubSection{Reference Elimination}
\label{sec:RefElim}

The reference elimination transformation is what enables the
alias-free memory model in the Move Prover, which is one of the most
important factors contributing to the speed and reliability of the
system.  In most software verification and static analysis systems,
the explosion in number of possible aliasing relationships between
references leads either to high computational complexity or harsh
approximations.

In Move, the reference system is based on
\emph{borrow semantics}~\cite{BORROW_SEM} as in the Rust programming language.
The initial borrow must come from either
a global memory or
a local variable on stack
(both referred to as \emph{locations} from now on).
For local variables, one can create
immutable references (with syntax |&x|) and
mutable references (with syntax |&mut x|).
For global memories, the references can be created via the
|borrow_global| and |borrow_global_mut| built-ins.
Given a reference to a whole struct,
field borrowing can occur via |&mut x.f| and |&x.f|.
Similarly, with a reference to a vector,
element borrowing occurs via native functions
|Vector::borrow(v, i)| and |Vector::borrow_mut(v, i)|.
Move provides the following guarantees,
which are enforced by the borrow checker:

\begin{itemize}
\item For any location, there can be
  either exactly one mutable reference,
  or $n$ immutable references.
  Enforcing this rule is similar to enforcing the borrow semantics
  in Rust, except for global memories, which do not exist in Rust.
  For global memories,
  this rule is enforced via the |acquires| annotations.
  Using Fig.~\ref{fig:AccountDef} as an example,
  function |withdraw| |acquires| the |Account| global location,
  therefore, |withdraw| is prohibited from calling any other function that might
  also borrow or modify the |Account| global memory (e.g., |deposit|).
\item The lifetime of references to data on the stack
  cannot exceed the lifetime of the stack location.
  This includes global memories borrowed inside a function as well---a
  reference to a global memory cannot be returned from the function,
  neither in whole nor in parts.
\end{itemize}

\noindent These properties effectively permit the \emph{elimination of
  references} from a Move program, eliminating need to reason about aliasing.

\Paragraph{Immutable References}

Immutable references are replaced by values.
An example of the applied transformation is shown below. We remove the reference
type constructor and all reference-taking operations from the code:

\begin{Move}
  fun select_f(s: &S): &T { &s.f } @\transform@ fun select_f(s: S): T { s.f }
\end{Move}

\noindent
When executing a Move program, immutable references are important to avoid copies
for performance and to enforce ownership; however, for symbolic reasoning on
correct Move programs, the distinction between immutable references and values
is unimportant.

\Paragraph{Mutable References}
\label{sec:RefElimMut}

Each mutation of a location |l| starts with an initial borrow for the whole data
stored in this location. This borrow creates a reference |r|.
As long as |r| is alive, Move code can either update its value (|*r = v|),
or derive a sub-reference (|r' = &mut r.f|).
The mutation ends when |r| (and the derived |r'|) go out of scope.

The borrow checker guarantees that during the mutation
of the data in |l|, no other reference can exist into the same data in |l|
-- meaning that it is impossible for other Move code to test whether the
value has mutated while the reference is held.

These semantics allow mutable references to be handled via
\emph{read-update-write} cycles.
One can create a copy of the data in |l| and
perform a sequence of mutation steps which are represented as purely functional
data updates.
Once the last reference for the data in |l| goes out of scope, the updated value
is written back to |l|.
This
converts an imperative program with references
into an imperative program which only has state updates on global memory or
variables on the stack,
with no aliasing.
We illustrate the basics of this approach by an example:

\begin{Move}
  fun increment(x: &mut u64) { *x = *x + 1 }
  fun increment_field(s: &mut S) { increment(&mut s.f) }
  fun caller(): S { let s = S{f:0}; update(&mut s); s }
  @\transform@
  fun increment(x: u64): u64 { x + 1 }
  fun increment_field(s: S): S { s[f = increment(s.f)] }
  fun caller(): S { let s = S{f:0}; s = update(s); s }
\end{Move}

\Paragraph{Dynamic Mutable References}

While the setup in above example covers a majority of the use cases in every day
Move code,
the general case is more complex, since the referenced location may not be known
statically.
Consider the following Move code:

\begin{Move}
  let r = if (p) &mut s1 else &mut s2;
  increment_field(r);
\end{Move}

\noindent Additional information in the logical encoding is required to deal
with such cases.
When a reference goes out of scope, we need to know
from which location it was derived in order to write back the updated value.
Fig.~\ref{fig:MutElim} illustrates the approach for doing this.
Essentially, a new type |Mut<T>|, which is internal to \MVP, is introduced to
track both the location from which |T| was derived and the value of |T|.
|Mut<T>| supports the following operations:

\begin{itemize}
\item |Mvp::mklocal(value, LOCAL_ID)| creates a new mutation value for a local
  with the given local id.  A local id uniquely identifies a local variable
  in the function.
\item Similarly, |Mvp::mkglobal(value, TYPE_ID, addr)| creates a new
  mutation for a global with given type and address.
\item With |r' = Mvp::field(r, FIELD_ID)| a mutation value for a sub-reference is
  created for the identified field.
\item The value of a mutation is replaced with |r' = Mvp::set(r, v)| and
  retrieved with |v = Mvp::get(r)|.
\item With the predicate |Mvp::is_local(r, LOCAL_ID)| one can test whether |r|
  was derived from the given local, and with |Mvp::is_global(r, TYPE_ID, addr)|
  for a specific global location.~%
  |Mvp::is_field(r, FIELD_ID)| tests whether |r| is derived from the given field.
\end{itemize}

\begin{figure}[t!]
  \caption{Elimination of Mutable References}
  \label{fig:MutElim}
  \centering
\begin{MoveBoxNumbered}
  fun increment(x: &mut u64) { *x = *x + 1 }
  fun increment_field(s: &mut S) {
    let r = if (s.f > 0) &mut s.f else &mut s.g;
    increment(r)
  }
  fun caller(p: bool): (S, S) {
    let s1 = S{f:0, g:0}; let s2 = S{f:1, g:1};
    let r = if (p) &mut s1 else &mut s2;
    increment_field(r);
    (s1, s2)
  }
  @\transform@
  fun increment(x: Mut<u64>): Mut<u64> { Mvp::set(x, Mvp::get(x) + 1) }
  fun increment_field(s: Mut<S>): Mut<S> {
    let r = if (s.f > 0) Mvp::field(s.f, S_F) else Mvp::field(s.g, S_G);
    r = increment(r);
    if (Mvp::is_field(r, S_F))
      s = Mvp::set(s, Mvp::get(s)[f = Mvp::get(r)]);
    if (Mvp::is_field(r, S_G))
      s = Mvp::set(s, Mvp::get(s)[g = Mvp::get(r)]);
    s
  }
  fun caller(p: bool): S {
    let s1 = S{f:0, g:0}; let s2 = S{f:1, g:1};
    let r = if (p) Mvp::mklocal(s1, CALLER_s1)
            else Mvp::mklocal(s2, CALLER_s2);
    r = increment_field(r);
    if (Mvp::is_local(r, CALLER_s1))
      s1 = Mvp::get(r); @\label{line:WriteBack}@
    if (Mvp::is_local(r, CALLER_s2))
      s2 = Mvp::get(r);
    (s1, s2)
  }
\end{MoveBoxNumbered}
\end{figure}

\MVP implements the illustrated transformation by construction a \emph{borrow
  graph} from the program via data flow analysis.  This graph tracks both when
references are released as well as how they relate to each other: e.g.~%
|r' = &mut r.f| creates an edge from |r| to |r'| labeled with |f|, and~%
|r' = &mut r.g| creates another also starting from |r|.  The borrow analysis is
inter-procedural, requiring computed summaries for the borrow graph of called
functions.

The resulting borrow graph is then used to guide the transformation, inserting
the operations of the |Mut<T>| type as illustrated in
Fig~\ref{fig:MutElim}. Specifically, when the borrow on a reference ends, the
associated mutation value must be written back to its parent mutation or the
original location (e.g. line~\ref{line:WriteBack} in
Fig.~\ref{fig:MutElim}). The presence of multiple possible origins leads to case
distinctions via |Mvp::is_X| predicates; however, these cases are rare in actual
Move programs.

\SubSection{Global Invariant Injection}
\label{sec:GlobalInvariants}

Correctness of smart contracts is largely about the correctness of the
blockchain state, so global invariants are particular important in the
move specification language.  For example, in the Diem framework,
global invariants can capture the requirement that an account be
accompanied by various other types that are be stored at the same
address and the requirement certain state changes are only permitted
for certain accounts by the access control scheme.

Most software verification tools prove that functions preserve
invariants by assuming the invariant at the entry to each function and
proving them at the exit.  In a module or class, it is only necessary
to prove that invariants are preserved by public functions, since
invariants are often violated internally in the implementation of a module or
class. An earlier version of the Move Prover used exactly this approach.

The current implementation of the Prover takes the opposite approach: it ensures
that invariants hold after every instruction, unless explicitly directed to
suspend some invariants by a user.  This \emph{fine-grained} approach has
performance advantages, because, unless suspended, \emph{invariants are only
  proven when an instruction is executed that could invalidate them}, and the
proofs are often computationally simple because \emph{the change from a single
  instruction is usually small}.  Relatively few invariants are suspended, and,
when they are, it is over a relatively small span of instructions, preserving
these advantages.  There is another important advantage, which is that
invariants hold almost everywhere in the code, so they are available to approve
other properties, such as abort conditions. For example, if a function accesses
type |T1| and then type |T2|, the access to |T2| will never abort if the
presence of |T1| implies the presence of |T2| at every state in the body of the
function.  This situation occurs with some frequency in the Diem framework.

\Paragraph{Invariant Types and Proof Methodology}

\emph{Inductive} invariants are properties declared in Move modules that must
(by default) hold for the global memory at all times. Those invariants often
quantify over addresses (See Fig.~\ref{fig:AccountSpec} for example.) Based on
Move's borrow semantics, inductive invariants don't need to hold while memory is
mutated because the changes are not visible to other code until the change is written back.
This is reflected by the reference elimination described in
Sec.~\ref{sec:RefElim},

\emph{Update} invariants are properties that relate two states, a previous state
and the current state.  Typically they are enforced after an update of global
memory. The |old| operator is used to evaluate specification expressions in the
previous state.

Verification of both kinds of invariants can be \emph{suspended}. That means,
instead of being verified at the time a memory update happens, they are verified
at the call site of the function which updates memory. This feature is
necessitated by fine-grained invariant checking, because invariants sometimes do
not hold in the midst of internal computations of a module. For example, a
relationship between state variables may not hold when the variables are being
updated sequentially.  Functions with external callers (public or script
functions) cannot suspend invariant verification, since the invariants are
assumed to hold at the beginning and end of each such function.

Inductive invariants are proven by induction over the evolution of the global
memory. The base case is that the invariant must hold in the empty state that
precedes the genesis transaction.  For the induction step, we can assume that
the invariant holds at each verified function entry point for which it is not
suspended, and now must prove that it holds after program points which are
either direct updates of global memory, or calls to functions which suspend
invariants.

For update invariants, no induction proof is needed, since they just relate two
memories.  The pre-state is some memory captured before an update happens, and
the post state the current state.

\Paragraph{Modular Verification}

We wish to support open systems to which untrusted modules can be added
with no chance of violating invariants that have already been proven. For each
invariant, there is a defined subset of Move modules (called a
\textit{cluster}). If the invariant is proven for the modules in the cluster, it
is guaranteed to hold in all other modules -- even those that were not yet
defined when the invariant was proven.  The cluster must contain every function
that can invalidate the invariant, and, in case of invariant suspension, all
callers of such a function.  Importantly, functions outside the cluster can
never invalidate an invariant. Those functions trivially preserve the
invariant, so it is only necessary to verify functions defined in the cluster.


\MVP verifies a given set of modules at a time (typically one).  The modules
being verified are called the \textit{target modules}, and the global invariants
to be verified are called \textit{target invariants}, which are all invariants
defined in the target modules. The cluster is then the smallest set as specified
above such that all target modules are contained.

\Paragraph{Basic Translation}

\begin{Figure}
  \caption{Basic Global Invariant Injection}
  \label{fig:GlobalInvariants}
  \centering
\begin{MoveBox}
  fun f(a: address) {
    let r = borrow_global_mut<S>(a);
    r.value = r.value + 1
  }
  invariant [I1] forall a: address: global<S>(a).value > 0;
  invariant [I2] update forall a: address:
      global<S>(a).value > old(global<S>(a).value);
  @\transform@
  fun f(a: address) {
    spec assume I1;
    Mvp::snapshot_state(I2_BEFORE);
    r = <increment mutation>;
    spec assert I1;
    spec assert I2[old = I2_BEFORE];
  }
\end{MoveBox}
\end{Figure}

We first look at injection of global invariants in the absence of
type parameters. Fig.~\ref{fig:GlobalInvariants} contains an
example for the supported invariant types and their injection into code. The
first invariant, |I1|, is an inductive invariant. It is assumed on function
entry, and asserted after the state update. The second, |I2|, is an update
invariant, which relates pre and post states. For this a state snapshot is
stored under some label |I2_BEFORE|, which is then used in an assertion.

Global invariant injection is optimized by knowledge of the prover, obtained by
static analysis, about accessed and modified memory.  Let |accessed(f)| be the
memory accessed by a function, and |modified(f)| be the memory modified. Let
|accessed(I)| by an invariant (including transitively by all functions it
calls).

\begin{itemize}
\item Inject |assume I| at entry to |f| \emph{if} |accessed(f)| has overlap with
  |accessed(I)|.
\item Inject |assert I| after each program step if one of the following is true
  (a) the step modifies a memory location |M in accessed(I)| or, (b) the step is
  a call to function |f'| in which |I| is suspended and |modifies(f')|
  intersects with |accessed(I)|.  Also, if |I| is an update invariant, inject a
  save of a memory snaptshot before the update or call.
\end{itemize}

\vspace{-1ex}
\Paragraph{Genericity}

\begin{Figure}
  \caption{Global Invariant Injection and Genericity}
  \label{fig:Genericity}
  \centering
\begin{MoveBox}
  invariant [I1] global<S<u64>>(0).value > 1;
  invariant<T> [I2] global<S<T>>(0).value > 0;
  fun f(a: address) { borrow_global_mut<S<u8>>(0).value = 2 }
  fun g<R>(a: address) { borrow_global_mut<S<R>>(0).value = 3 }
  @\transform@
  fun f(a: address) {
    spec assume I2[T = u8];
    <<mutate>>
    spec assert I2[T = u8];
  }
  fun g<R>(a: address) {
    spec assume I1; spec assume I2[T = R];
    <<mutate>>
    spec assert I1; spec assert I2[T = R];
  }
\end{MoveBox}
\end{Figure}

Generic type parameters make the problem of determining whether a function can
modify an invariant more difficult.  Consider the example in
Fig.~\ref{fig:Genericity}. Invariant |I1| holds for a specific type
instantiation |S<u64>|, whereas |I2| is generic over all type instantiations for
|S<T>|.

The non-generic function |f| which works on the instantiation |S<u8>| will have
to inject the \emph{specialized} instance |I2[T = u8]|. The invariant |I1|,
however, does not apply for this function, because there is no overlap with
|S<u64>|.  In contrast, |g| is generic in type |R|, which could be instantiated
to |u64|. So, |I1|, which applies to |S<u64>| needs to be injected
in addition to |I2|.

The general solution depends on type unification.  Given the accessed memory of
a function |f<R>| and an invariant |I<T>|, we compute the pairwise unification
of memory types. Those types are parameterized over |R| resp. |T|. Successful
unification results in a substitution for both type parameters, and we include
the invariant with |T| specialized according to the substitution.

\SubSection{Monomorphization}
\label{sec:Mono}

Monomorphization is a transformation which removes generic types from a Move
program by \emph{specializing} the program for relevant type instantiations.  In
the context of verification, the goal is that the specialized program verifies
if and only if the generic program verifies in an encoding which supports types
as first class values. We expect the specialized program to verify faster
because it avoids the problem of generic representation of values, supporting
a multi-sorted representation in the SMT logic.

\begin{Figure}
\caption{Basic Monomorphization}
\label{fig:Mono}
\centering
\begin{MoveBox}
  struct S<T> { .. }
  fun f<T>(x: T) { g<S<T>>(S(x)) }
  fun g<S:key>(s: S) { move_to<S>(.., s) }
  @\transform@
  struct T{}
  struct S_T{ .. }
  fun f_T(x: T) { g_S_T(S_T(x)) }
  fun g_S_T(s: S_T) { move_to<S_T>(.., s) }
\end{MoveBox}
\end{Figure}

To verify a generic function for all possible instantiations, monomorphization
skolemizes the type parameter, i.e.~the function is verified for a new type
with no special properties that represents an arbitrary type.  It then
specializes all called functions and used data types with this new type and any
other concrete types they may use.  Fig.~\ref{fig:Mono} sketches this approach.

However, this approach has one issue: the type of genericity Move provides does
not allow for full type erasure (unlike many programming languages) because
types are used to \emph{index} global memory (e.g. |global<S<T>>(addr)| where
|T| is a generic type). Consider the following Move function:

\begin{Move}
  fun f<T>(..) { move_to<S<T>>(s, ..); move_to<S<u64>>(s, ..) }
\end{Move}

\noindent Depending on how |T| is instantiated, this function behaves
differently.  Specifically, if |T| is instantiated with |u64| the function will
always abort at the second |move_to|, since the target location is already
occupied.

The important property enabling monomorphization in the presence of such type
dependent code is that one can identify the situation by looking at the memory
accessed by code and injected specifications. From this one can derive
\emph{additional instantiations of the function} which need to be verified. In
the example above, verifying both |f_T| and an instantiation |f_u64| will cover
all relevant cases of the function behavior.

The algorithm for computing the instances that require verification works as
follows. Let |f<T1,..,Tn>| be a verified target function which has all
specifications injected and inlined function calls expanded.
\begin{itemize}
\item For each memory~~|M in modified(f)|, if there is a memory~%
  |M' in modified(f) + accessed(f)| such that |M| and |M'| can unify via
  |T1,..,Tn|, collect an instantiation of the type parameters |Ti| from the
  resulting substitution. This instantiation may not assign values to all type
  parameters, and those unassigned parameters stay as is. For instance,~%
  |f<T1, T2>| might have a partial instantiation |f<T1, u8>|.
\item Once all partial instantiations are computed, the set is
  extended by unifying the instantiations against each other. If |<T>| and
  |<T'>| are in the set, and they unify under the substitution |s|, then
  |<s(T)>| will also be part of the set.  For example, consider |f<T1, T2>|
  which modifies |M<T1>| and |R<T2>|, as well as accesses |M<u64>| and
  |R<u8>|. From this the instantiations |<u64, T2>| and |<T1, u8>| are computed,
  and the additional instantiation |<u64, u8>| will be added to the set.
\item If after computing and extending instantiations any type parameters
  remain, they are skolemized into a given type as described earlier.
\end{itemize}

\noindent To understand the correctness of this procedure, consider the
following arguments (a full formal proof is outstanding):

\begin{itemize}
\item \emph{Direct interaction} Whenever a modified memory |M<t>| can influence
  the interpretation of |M<t'>|, a unifier must exist for the types |t| and |t'|,
  and an instantiation will be verified which covers the overlap of |t| and
  |t'|.
\item \emph{Indirect interaction} If there is an overlap between two types
  which influences whether another overlap is semantically relevant, the
  combination of both overlaps will be verified via the extension step.
\end{itemize}

Notice that even though it is not common in regular Move code to work with both
memory |S<T>| and, say, |S<u64>| in one function, there is a scenario where such
code is implicitly created by injection of global invariants. Consider the
example in Fig.~\ref{fig:Genericity}. The invariant |I1| which works on |S<u64>|
is injected into the function |g<R>| which works on |S<R>|. When monomorphizing
|g|, we need to verify an instance |g_u64| in order to ensure that |I1| holds.


\Section{Analysis}

\Paragraph{Reliability and Performance}

The three improvements described above resulted in a major qualitative
change in performance and reliability. In the version of \MVP released
in September 2020, correct examples verified fairly quickly and
reliably.  But that is because we needed speed and reliability, so we
disabled some properties that always timed out and others that timed
out unpredictably when there were small changes in the framework.  We
learned that incorrect programs or specifications would time out
predictably enough that it was a good bet that examples that timed out
were erroneous.  However, localizing the error to fix it was
\emph{very} hard, because debugging is based on a counterexample that
violates the property, and getting a counterexample requires
termination!

With each of the transformations described, we witnessed significant speedups
and, more importantly, reductions in timeouts.  Monomorphization was the last
feature implemented, and, with it, timeouts almost disappeared. Although this
was the most important improvement in practice, it is difficult to quantify
because there have been many changes in Diem framework, its specifications,
\MVP, and even the Move language over that time.

It is simpler (but less important) to quantify the changes in run time of \MVP
on one of our more challenging modules, the |DiemAccount| module, which is the
biggest module in the Diem framework. This module implements basic functionality
to create and maintain multiple types of accounts on the blockchain, as well as
manage their coin balances. It was called |LibraAccount| in release 1.0 of \MVP,
and is called |DiemAccount| today. The comparison requires various patches as
described in~\cite{MVP_ARTIFACT}. The table below lists the consolidated numbers
of lines, functions, invariants, conditions (requires, ensures, and aborts-if),
as well as the verification times:

{
\setlength{\tabcolsep}{6pt}
\vspace{2ex}
\begin{tabular*}{0.9\textwidth}{cccccc}
  \hline
  \hline
  Module & Lines & Functions & Invariants & Conditions & Timing \\
  \hline
  LibraAccount & 1975 & 72 & 10 & 113 & \textbf{9.899s} \\
  DiemAccount & 2554 & 64 & 32 & 171 & \textbf{7.340s} \\
  \hline
\end{tabular*}
\vspace{2ex}
}

\noindent Notice that |DiemAccount| has significantly grown in size compared to
the older version. Specifically, additional specifications have been
added. Moreover, in the original |LibraAccount|, some of the most complex
functions had to be disabled for verification because the old version of \MVP
would time out on them. In contrast, in |DiemAccount| and with the new version,
all functions are verified. Verification time has been improved by roughly 20\%,
\emph{in the presence of three times more global invariants, and 50\% more function
conditions}.

We were able to observe similar improvements for the remaining of the 40 modules
of the Diem framework. All of the roughly half-dozen timeouts resolved after
introduction of the transformations described in this paper.



\Paragraph{Causes for the Improvements}

It's difficult to pin down and measure exactly why the three transformations
described improved performance and reliability so dramatically.  We have
explained some reasons in the subsections above: the alias-free memory model
reduced search through combinatorial sharing arrangments, and the fine-grained
invariant checking results in simpler formulas for the SMT solver.

We found that most timeouts in specifications stemmed from our liberal use of
quantifiers.  To disprove a property $P_0$ after assuming a list of properties,
$P_1, \ldots p_n$, the SMT solver must show that
$\neg P_0 \wedge P_1 \wedge \ldots \wedge P_n$ is satisfiable.  The search
usually involves instantiating universal quantifiers in $P_1, \ldots, P_n$.  The
SMT solver can do this endlessly, resulting in a timeout. Indeed, we often found
that proving a post-condition false would time out, because the SMT solver was
instantiating quantifiers to find a satisfying assignment of
$P_1 \wedge \ldots \wedge P_n$.  Simpler formulas result in fewer intermediate
terms during solving, resulting in fewer opportunities to instantiate quantified
formulas.

We believe that one of the biggest impacts, specifically on
removing timeouts and improving predictability, is monomorphization. The
reason for this is that monomorphization allows a multi-sorted representation
of values in Boogie (and eventually the SMT solver). In contrast, before
monomorphization, we used a universal domain for values in order to represent
values in generic functions, roughly as follows:

\begin{Move}
  type Value = Num(int) | Address(int) | Struct(Vector<Value>) | ...
\end{Move}

\noindent This creates a large overhead for the SMT solver, as we need to
exhaustively inject type assumptions (e.g. that a |Value| is actually an
|Address|), and pack/unpack values. Consider a quantifier like~%
|forall a: address: P(x)| in Move. Before monomorphization, we have to represent
this in Boogie as~%
|forall a: Value: is#Address(a) => P(v#Address(a))|. This quantifier is
triggered where ever |is#Address(a)| is present, independent of the structure of
|P|. Over-triggering or inadequate triggering of quantifiers is one of the
suspected sources of timeouts, as also discussed in~\cite{BUTTERFLY}.

Moreover, before monomorphization, global memory was indexed in Boogie by an
address and a type instantiation. That is, for |struct R<T>| we would
have one Boogie array |[Type, int]Value|. With monomorphization, the type index
is eliminated, as we create different memory variables for each type
instantiation.  Quantification over memory content works now on a one-dimensional
instead of an n-dimensional Boogie array.

\Paragraph{Discussion and Related Work}

Many approaches have been applied to the verification of smart contracts; see
e.g. the
surveys~\cite{liu2019survey,CONTRACT_VERIFICATION}. \cite{CONTRACT_VERIFICATION}
refers to at least two dozen systems for smart contract verification. It
distinguishes between \emph{contract} and \emph{program} level approaches. Our
approach has aspects of both: we address program level properties via pre/post
conditions, and contract (``blockchain state'') level properties via global
invariants. To the best of our knowledge, among the existing approaches, the
Move ecosystem is the first one where contract programming and specification
language are fully integrated, and the language is designed from first
principles influenced by verification. Methodologically, Move and the Move
prover are thereby closer to systems like Dafny~\cite{DAFNY}, or the older
Spec\# system~\cite{SPECSHARP}, where instead of adding a specification approach
posterior to an existing language, it is part from the beginning. This allows us
not only to deliver a more consistent user experience, but also to make
verification technically easier by curating the programming language.


In contrast to other approaches that only focus on specific vulnerability
patterns~\cite{mythril,oyente,maian,securify}, \MVP offers a universal
specification language. To the best of our knowledge, no existing specification
approach for smart contracts based on inductive Hoare logic has similar
expressiveness. We support universal quantification over arbitrary memory
content, a suspension mechanism of invariants to allow non-atomic construction
of memory content, and generic invariants.  For comparison, the SMT Checker
build into \solidity \cite{solidity,solcverify,DBLP:conf/esop/HajduJ20}
does not support quantifiers, because it interprets programming language
constructs (requires and assert statements) as specifications and has no
dedicated specification language. While in Solidity one can simulate aspects of
global invariants using modifiers by attaching pre/post conditions, this is not
the same as our invariants, which are guaranteed to hold independent of whether
a user may or (accidentally) may not attach a modifier, and which are optimized
to be only evaluated as needed.

While the expressiveness of Move specifications comes with the price of
undecidability and the dependency from heuristics in SMT solvers, \MVP deals
with this by its elaborated translation to SMT logic, as described in this
paper. The result is a practical verification system that is fully integrated
into the Diem blockchain production process, running in continuous integration,
which is (to the best of our knowledge) a first in the industry.

The individual techniques we described are novel each by
themselves. \emph{Reference elimination} relies on borrow semantics, similar as
in the Rust~\cite{rust} language.  We expect reference elimination to apply for
the safe subset of Rust, though some extra work would be needed to deal with
references aggregated by structs.  However, we are not aware of that something
similar has been attempted in existing Rust verification work
\cite{prusti,smack,nopanic,crust}. \emph{Global invariant injection} and the
approach to minimize the number of assumptions and assertions is not applied in
any existing verification approach we know of; however, we co-authored a while
ago a similar line of work for \emph{runtime checking} of invariants in
Spec\#~\cite{StateConstraintsPatent}, yet that work never left the conceptual
state. \emph{Monomorphization} is well known as a technique for compiling
languages like C++ or Rust, where it is called specialization; however, we are
not aware of it being generalized for modular verification of generic code where
full type erasure is not possible, as it is the case in Move.

\Paragraph{Future Work}

\MVP is conceived as a tool for achieving higher assurance systems, not as a bug
hunting tool. Having at least temporarily achieved satisfactory performance and
reliability, we are turning our attention to the question of the goal of higher
assurance, which raises several issues.  If we're striving for high assurance,
it would be great to be able to measure progress towards that goal.  Since
system requirements often stem from external business and regulatory needs,
lightweight processes for exposing those requirements so we know what needs to
be formally specified would be highly desirable.

As with many other systems, it is too hard to write high-quality specifications.
Our current specifications are more verbose than they need to be, and we are
working to require less detailed specifications, especially for individual
functions.  We could expand the usefulness of \MVP for programmers if we could
make it possible for them to derive value from simple reusable specifications.
Finally, software tools for assessing the consistency and completeness of formal
specifications would reduce the risk of missing bugs because of specification
errors.

However, as more complex smart contracts are written and as more people write
specifications, we expect that the inherent computational difficulty of solving
logic problems will reappear, and there will be more opportunities for improving
performance and reliability.  In addition to translation techniques, it will be
necessary to identify opportunities to improve SMT solvers for the particular
kinds of problems we generate.


\Section{Conclusion}

We described key aspects of the Move prover (\MVP), a tool for formal
verification of smart contracts written in the Move language. \MVP has been
successfully used to verify large parts of the Diem framework, and is used in
continuous integration in production. The specification language is expressive,
specifically by the powerful concept of global invariants.  We described key
implementation techniques which (as confirmed by our benchmarks) contributed to
the scalability of \MVP. One of the main areas of our future research is to
improve specification productivity and reduce the effort of reading and writing
specs, as well as to continue to improve speed and predictability of
verification.


\Paragraph{Acknowledgements}

This work would not have been possible without the many contributions of the
Move platform team and collaborators.  We specifically like to thank Bob Wilson,
Clark Barrett, Dario Russi, Jack Moffitt, Jake Silverman, Mathieu Baudet,
Runtian Zhou, Sam Blackshear, Tim Zakian, Todd Nowacki, Victor Gao, and Kevin
Cheang.

\appendix
\newpage
\Section{Injection of Function and Data Specifications}

In this appendix we describe, for the interested reader, the design of function
and data specification injection (requires, ensures, aborts\_if, modifies,
emits, and data invariants).  While the impact on speed and reliability of
verification might have been not that significant, these designs were fine tuned
over time as well. With this, a comprehensive coverage of the translation of all
specification constructs in \MVP is provided.

\SubSection{Function Specifications}

The injection of basic function specifications is illustrated in
Fig.~\ref{fig:RequiresEnsuresAbortsIf}.  An extension of the Move source
language is used to specify abort behavior. With~%
|fun f() { .. } onabort { conditions }| a Move function is defined where
|conditions| are assume or assert statements that are evaluated at every program
point the function aborts (either implicitly or with an |abort| statement). This
construct simplifies the presentation and corresponds to a per-function abort
block on bytecode level which is target of branching.

\begin{Figure}
  \caption{Requires, Ensures, and AbortsIf Injection}
  \label{fig:RequiresEnsuresAbortsIf}
  \centering
\begin{MoveBoxNumbered}
  fun f(x: u64, y: u64): u64 { x + y }
  spec f {
    requires x < y;
    aborts_if x + y > MAX_U64;
    ensures result == x + y;
  }
  fun g(x: u64): u64 { f(x, x + 1) }
  spec g {
    ensures result > x;
  }
  @\transform@
  fun f(x: u64, y: u64): u64 {
    spec assume x < y;
    let result = x + y;
    spec assert result == x + y;     // ensures of f
    spec assert                      // negated abort_if of f
      !(x + y > MAX_U64); @\label{line:aborts_holds_not}@
    result
  } onabort {
    spec assert                      // abort_if of f
      x + y > MAX_U64; @\label{line:aborts_holds}@
  }
  fun g(x: u64): u64 {
    spec assert x < x + 1;           // requires of f
@$\textrm{\it if inlined}$\label{line:inline}@
    let result = inline f(x, x + 1);
@$\textrm{\it elif opaque}$\label{line:opaque}@
    if (x + x + 1 > MAX_U64) abort;  // aborts_if of f
    spec assume result == x + x + 1; // ensures of f
@$\textrm{\it endif}$@
    spec assert result > x;          // ensures of g
    result
  }
\end{MoveBoxNumbered}
\end{Figure}

An aborts condition is translated into two different asserts: one where the
function aborts and the condition must hold (line~\ref{line:aborts_holds}), and
one where it returns and the condition must \emph{not} hold
(line~\ref{line:aborts_holds_not}). If there are multiple |aborts_if|, they are
or-ed. If there is no abort condition, no asserts are generated. This means
that once a user specifies aborts conditions, they must completely cover the
abort behavior of the code. (The prover also provides an option to relax this
behavior, where aborts conditions can be partial and are only enforced on
function return.)

For a function call site we distinguish two variants: the call is \emph{inlined}
(line~\ref{line:inline}) or it is \emph{opaque} (line~\ref{line:opaque}).  For
inlined calls, the function definition, with all injected assumptions and
assertions turned into assumptions (as those are considered proven) is
substituted. For opaque functions the specification conditions are inserted as
assumptions. Methodologically, opaque functions need precise specifications
relative to a particular objective, where as in the case of inlined functions
the code is still the source of truth and specifications can be partial or
omitted. However, inlining does not scale arbitrarily, and can be only used for
small function systems.

Notice we have not discussed the way how to deal with relating pre and post
states yet, which requires taking snapshots of state (e.g.~%
|ensures x == old(x) + 1|); the example in
Fig.~\ref{fig:RequiresEnsuresAbortsIf} does not need it. Snapshots of state
are discussed for global update invariants in Sec.~\ref{sec:GlobalInvariants}.

\Paragraph{Modifies}

\begin{Figure}
  \caption{Modifies Injection}
  \label{fig:Modifies}
  \centering
\begin{MoveBoxNumbered}
  fun f(addr: address) { move_to<T>(addr, T{}) }
  spec f {
    pragma opaque;
    ensures exists<T>(addr);
    modifies global<T>(addr);
  }
  fun g() { f(0x1) }
  spec g {
    modifies global<T>(0x1); modifies global<T>(0x2);
  }
  @\transform@
  fun f(addr: address) {
    let can_modify_T = {addr};      // modifies of f
    spec assert addr in can_modify; // permission check @%
                                            \label{line:modifies_permission}@
    move_to<T>(addr, T{});
  }
  fun g() {
    let can_modify_T = {0x1, 0x2};  // modifies of g
    spec assert {0x1} <= can_modify_T; // permission check @%
                                            \label{line:modifies_call_permission}@
    spec havoc global<T>(0x1);      // havoc modified memory @%
                                            \label{line:modifies_havoc}@
    spec assume exists<T>(0x1);     // ensures of f
  }
\end{MoveBoxNumbered}
\end{Figure}

The |modifies| condition specifies that a function only changes specific memory.
It comes in the form |modifies global<T>(addr)|, and its injection is
illustrated in Fig.~\ref{fig:Modifies}.

A type check is used to ensure that if a function has one or more~%
|modifies| conditions all called functions which are \emph{opaque} have a
matching modifies declaration. This is important so we can relate the callees
memory modifications to that what is allowed at caller side.

At verification time, when an operation is performed which modifies memory, an
assertion is emitted that modification is allowed
(e.g. line~\ref{line:modifies_permission}). The permitted addresses derived from
the modifies clause are stored in a set |can_modify_T| generated by the
transformation. Instructions which modify memory are either primitives (like
|move_to| in the example) or function calls. If the function call is inlined,
modifies injection proceeds (conceptually) with the inlined body. For opaque
function calls, the static analysis has ensured that the target has a modifies
clause.  This clause is used to derive the modified memory, which must be a
subset of the modified memory of the caller
(line~\ref{line:modifies_call_permission}).

For opaque calls, we also need to \emph{havoc} the memory they modify
(line~\ref{line:modifies_havoc}), by which is meant assigning an unconstrained
value to it. If present, |ensures| from the called function, injected as
subsequent assumptions, are further constraining the modified memory.

\Paragraph{Emits}

\begin{Figure}
  \caption{Emits Injection}
  \label{fig:Emits}
  \centering
\begin{MoveBoxNumbered}
  use Std::Event;
  struct E has drop, store { m: u64 }
  fun f(h: &mut Event::EventHandle<E>, x: u64) {
    Event::emit_event(h, E{m:0}); @\label{line:emit_event}@
    if (x > 0) {
      Event::emit_event(h, E{m:x});
    }
  }
  spec f {
    emits E{m:0} to h; @\label{line:emits}@
    emits E{m:x} to h if x > 0; @\label{line:emits_if}@
  }
  @\transform@
  fun f(h: &mut Event::EventHandle<E>, x: u64) {
    es = Mvp::ExtendEventStore(es, h, E{m:0}); // emitting event @\label{line:extend_es}@
    if (x > 0) {
      es = Mvp::ExtendEventStore(es, h, E{m:x}); // emitting event
    }
    let actual_es = Mvp::subtract(es, old(es)); // events emitted by f @\label{line:actual_es}@
    let expected_es = Mvp::CondExtendEventStore( // specified events @\label{line:expected_es}@
      Mvp::ExtendEventStore(Mvp::EmptyEventStore, E{m:x}, h),
      E{m:x}, h, x>0);
    spec assert Mvp::includes(expected_es, actual_es); // spec completeness @\label{line:emits_completeness}@
    spec assert Mvp::includes(actual_es, expected_es); // spec relevance @\label{line:emits_relevance}@
  }
\end{MoveBoxNumbered}
\end{Figure}

The injection for the |emits| clause is illustrated in Fig.~\ref{fig:Emits}. The
|emits| clause specifies the events that a function is expected to emit. It
comes in the form |emits message to handle if condition| (e.g.,
line~\ref{line:emits_if}). The condition part (i.e., |if condition|) can be
omitted if the event is expected to be emitted unconditionally (e.g.,
line~\ref{line:emits}).

The function call to |Event::emit_event| (e.g., line~\ref{line:emit_event}) is
transformed into the statement to extend |es| with the event to emit (e.g.,
line~\ref{line:extend_es}). |es| is a global variable of type |EventStore| which
is a map where the key is an event handle and the value is the event stream of
the handle (modeled as a bag of messages).

In line~\ref{line:actual_es}, |actual_es| represents the portion of the
EventStore that only comprises the events that the program (i.e., |f|) actually
emits. In line~\ref{line:expected_es}, |expected_es| is constructed from the
|emits| specification which contains all of the expected events specified by the
|emits| clauses. Having these, two assertions using |Mvp::includes| (multiset
inclusion relation per event handle) are injected.

\begin{itemize}
\item One asserts that |expected_es| includes |actual_es|, meaning that the
  function only emits the events that are expected (e.g.,
  line~\ref{line:emits_completeness}). This would be violated if there is any
  event emitted by |f| that is not covered by some |emits| clause.
\item The other asserts that |actual_es| includes |expected_es|, meaning that
  the function emits all of the events that are expected (e.g.,
  line~\ref{line:emits_relevance}). This would be violated if |f| does not emit
  the expected event which a |emits| clause specifies.
\end{itemize}

We also handle opaque calls properly although it is not illustrated in
Fig.~\ref{fig:Emits}. Suppose |f| is an opaque function, and another function
|g| calls |f|. In the transformation of |g|, the event store |es| extends with
the expected events of |f| (i.e., the events specified by the |emits| clauses of
|f|) in a similar way to how |expected_es| is constructed in
line~\ref{line:expected_es}.

\vspace{-2ex}
\SubSection{Data Invariants}

\begin{Figure}
  \caption{Data Invariant Injection}
  \label{fig:DataInvariants}
  \centering
\begin{MoveBoxNumbered}
  struct S { a: u64, b: u64 }
  spec S { invariant a < b }
  fun f(s: S): S {
    let r = &mut s;
    r.a = r.a + 1;
    r.b = r.b + 1;
    s
  }
  @\transform@
  fun f(s: S): S {
    spec assume s.a < s.b;      // assume invariant for s
    let r = Mvp::local(s, F_s); // begin mutation of s
    r = Mvp::set(r, Mvp::get(r)[a = Mvp::get(r).a + 1]);
    r = Mvp::set(r, Mvp::get(r)[b = Mvp::get(r).b + 1]);
    spec assert                 // invariant enforced
      Mvp::get(r).a < Mvp::get(r).b;
    s = Mvp::get(r);            // write back to s
    s
  }
\end{MoveBoxNumbered}
\end{Figure}

A data invariant specifies a constraint over a struct value. The value is
guaranteed to satisfy this constraint at any time. Thus, when a value is
constructed, the data invariant needs to be verified, and when it is consumed,
it can be assumed to hold.

In Move's reference semantics, construction of struct values is often done via a
sequence of mutations via mutable references. It is desirable that \emph{during}
such mutations, assertion of the data invariant is suspended. This allows to
state invariants which reference multiple fields, where the fields are updated
step-by-step.  Move's borrow semantics and concept of mutations provides a
natural way how to defer invariant evaluation: at the point a mutable reference
is released, mutation ends, and the data invariant can be enforced.  In other
specification formalisms, we would need a special language construct for
invariant suspension. Fig.~\ref{fig:DataInvariants} gives an example, and shows
how data invariants are reduced to assert/assume statements.

The implementation of data invariants hooks into reference elimination,
described in Sec.~\ref{sec:RefElim}. As part of this the lifetime of references
is computed. Whenever a reference is released and the mutated value is written
back, we also assert the data invariant. In addition, the data invariant is
asserted when a struct value is directly constructed.


\Section{Corrected Account Example}
\label{sec:CorrectedAccount}

To fix the verification errors from the account example in  Fig.~\ref{fig:AccountDef}
and Fig.~\ref{fig:AccountSpec}, the following changes would need to be made:

\begin{MoveBox}
module Account {
  ...

  fun withdraw(account: address, amount: u64) acquires Account {
    assert(amount <= AccountLimits::max_decrease(), Errors::invalid_argument()); // MISSING
    let balance = &mut borrow_global_mut<Account>(account).balance;
    assert(*balance >= amount, Errors::limit_exceeded());
    assert(*balance - amount >= AccountLimits::min_balance(), Errors::invalid_argument()); // MISSING
    *balance = *balance - amount;
  }

  spec transfer {
    ...
    aborts_if !exists<Account>(from_addr); // MISSING
    aborts_if !exists<Account>(to); // MISSING
    aborts_if amount > AccountLimits::max_decrease(); // MISSING
    aborts_if bal(from_addr) - amount < AccountLimits::min_balance(); // MISSING
  }
}
\end{MoveBox}


\newpage
\bibliographystyle{splncs04}
\bibliography{biblio}

\vfill

{\small\medskip\noindent{\bf Open Access} This chapter is licensed under the terms of the Creative Commons\break Attribution 4.0 International License (\url{http://creativecommons.org/licenses/by/4.0/}), which permits use, sharing, adaptation, distribution and reproduction in any medium or format, as long as you give appropriate credit to the original author(s) and the source, provide a link to the Creative Commons license and indicate if changes were made.}

{\small \spaceskip .28em plus .1em minus .1em The images or other third party material in this chapter are included in the chapter's Creative Commons license, unless indicated otherwise in a credit line to the material.~If material is not included in the chapter's Creative Commons license and your intended\break use is not permitted by statutory regulation or exceeds the permitted use, you will need to obtain permission directly from the copyright holder.}

\medskip\noindent\includegraphics{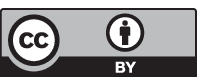}

\end{document}